# Can repulsion be induced by attraction: a role of ephrin-B1 in retinotectal mapping?

## Dmitry N. Tsigankov and Alexei A. Koulakov
Cold Spring Harbor Labaratory, Cold Spring Harbor, NY 11724, USA

We study a role of EphB receptors and their ligand ephrin-B1 in dorsal-ventral retinotopic mapping. Earlier studies suggested that ephrin-B1 acts as an attractant for EphB expressing axons. We address the results of the recent experiment in chick tectum (McLaughlin et al., 2003b) in which axons of retinal ganglion cells were shown to be repelled by high ephrin-B1 density. Thus it was proposed that ephrin-B1 might act as both attractant and repellent. We show that the same axonal behavior may follow from attraction to ephrin-B1 density and axonal competition for space. Therefore, we show how apparent repulsive interaction can be induced by a combination of attraction to the target and competitive interactions between axons. We suggest an experimental test that may distinguish repulsive interaction with the target from repulsion induced by attraction and competition.

## *Introduction*

Axons of retinal ganglion cells (RGC) projecting to optic tectum establish an orderly arrangement of connections. The spatial ordering of axonal termination points in tectum forms a basis for a topographic map that represents the visual world. This mapping occurs along two approximately orthogonal axes in tectum. The temporal-nasal (T-N) axis of the retina is mapped onto the anterior-posterior (A-P) tectal axis, and the dorsal-ventral (D-V) retinal axis along the lateral-medial (L-M) tectal axis.

Studies of the molecular mechanisms responsible for topographic mapping of D-V retinal axis revealed a key role of the EphB receptor tyrosine kinases and their ephrin-B ligands in map formation. EphB2, EphB3 and EphB4 are expressed in a low to high D-V gradient in RGC in chicks and mice (Holash and Pasquale, 1995; Birgbauer et al., 2000; Hindges et al., 2002; Mann et al., 2002) and ephrin-B1 is expressed in a low to high L-M gradient in chick tectum and mouse superior colliculus (Braisted et al., 1997; Hindges et al., 2002). Since the axons with high/low level of EphB expression project to the regions with high/low ephrin-B1 density, this implies an attractive interaction between EphB receptors and their ligands (O'Leary and Wilkinson, 1999). The attraction hypothesis is supported by experiments in chick (Holash and Pasquale, 1995), xenopus (Mann et al., 2002) and mice (Hindges et al., 2002). Thus, it is suggested that attractive interactions between EphB receptors and ephrin-B1 regulates mapping of RGC axons along D-V axis.

Nevertheless, mechanisms of the map formation in D-V direction are far from being clear. If there were only attractive interaction, all RGC axons would tend to project to the regions of high ephrin-B1 density and would preferentially extend their branches up the ephrin-B1 gradient. However, during map development, primary RGC axons lack D-V ordering and extend their branches toward their termination zones (TZs) either up or down the ephrin-B1 gradient (Simon and O'Leary, 1992; Yates et al., 2001). Thus, one type of interaction can hardly account for correct topographic mapping (Flanagan and Vanderhaeghen, 1998; Feldheim et al., 2000). A second interaction (repulsive or attractive) that opposes the attraction to ephrin-B1 density is needed to explain the map formation.

A recent experimental study reports that ephrin-B1 itself may provide the second type of interaction (McLaughlin et al., 2003b). It is shown that low ephrin-B1 density attracts RGC axon branches, while high density repels them. It is proposed that at the location of TZ two interactions cancel each other out thus determining the position of axonal TZ. To demonstrate the bifunctional action of ephrin-B1 as a repellent and an attractant, the regions of high ephrin-B1 density are artificially created by retroviral transfection in the chick tectum. Within these regions all branch extensions are redirected laterally (down the ephrin-B1 gradient, see Fig.1), thus being repelled by the high ephrin-B1 density. In addition, it is found that when the location of a TZ is coincident with the region of high ephrin-B1 density, this domain inhibits arborization and shapes the distribution of arbors. The repulsion of RGC axon branches from the regions of high ephrin-B1 density suggests a mechanism of the map formation and, seemingly, resolves the discussion of the source of the second interaction that results in the branch extensions down the ephrin-B1 gradient.

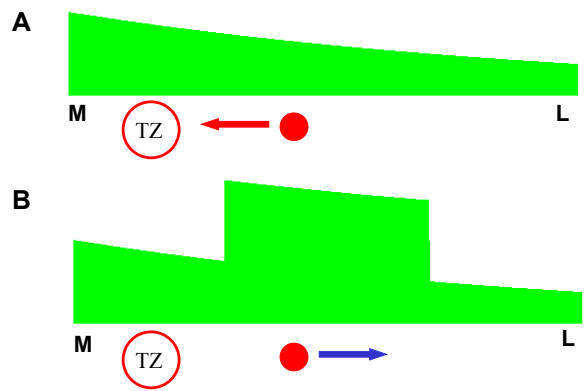

Figure 1 - Axonal repulsion from the regions of high ephrin-B density.
The directions of branch extensions are shown schematically for wild-type (A) and transfected (B) tectum as observed in (McLaughlin et al., 2003b). The profile of the ephrin-B1 density in tectum is shown in green. The axon (red dot) that have termination zone (red empty circle) medial from it, extends its branches up the ephrin-B1 gradient (red arrow) in wild-type tectum (A), redirects its branch extension down the gradient (blue arrow) within the region of high ephrin-B1 density (B). (M) medial, (L) lateral.

In this study we show that the same axonal behavior may result from a different mechanism based on axonal competition. It was proposed (Prestige and Willshaw, 1975; Fraser and Hunt, 1980; Feldheim et al., 2000) that axonal competition for space may provide the second interaction needed for map formation. Axonal competition leads to repulsion between branches of different axons. Repulsion from branches with higher EphB density may force axons down the ephrin-B1 gradient. We call this phenomenon a 'subway effect', since it is similar to behavior of passengers in the subway during rush hour. Note that our model relies on only one type of interaction (attractive) between RGC axons and target cells and axonal repulsion/competition. This is in contrast with other models which employ two types of interaction (e.g. attractive and repulsive) (McLaughlin et al., 2003a). We suggest an additional experimental test that may distinguish these two types of models.

## Model

To model topographic map development, we consider a square 100x100 array of RGC projecting to a 100x100 array of termination sites in tectum. Any terminal site in tectum can be occupied by one and only one RGC axon, which reflects competition between RGC axons. Each RGC axon is characterized by the expression levels of EphA and EphB receptors. Each termination site is described by the concentration of ephrinA and ephrinB. In wild type conditions, all density profiles have the same exponential form $\rho(n)=\exp(-n/100)$, where $n=1:100$ is the site position in tectum along the P-A axis for ephrinA and along the M-L axis for ephrinB, and it is the retinal position of RGC axons along the T-N axis for EphA and the V-D axis for EphB density profiles. In the transfected tectum, ephrinB density is increased by a constant level within the ectopic region. The discussion of the choice of concentration profiles can be found in (Koulakov and Tsigankov, 2003).

RGC projections initially form a random map. During development the map undergoes a series of reconstructions similar to what was used by other groups (Hope et al., 1976; Overton and Arbib, 1982). In this study the rules of the reconstruction reflect the chemoaffinity principle and are defined by the interactions between the chemical labels of RGC axons and the target cells. At each consecutive step of reconstruction we randomly choose a neighboring pair of axons in either A-P or L-M direction and exchange their positions with probabilities given by

$$P_{EXCHANGE} = \frac{1}{2} + \frac{1}{2}\tanh\left(2\alpha\left[RA(1)-RA(2)\right]\left[LA(1)-LA(2)\right]\right) \quad (1)$$

for the A-P direction, and

$$P_{EXCHANGE} = \frac{1}{2} - \frac{1}{2}\tanh\left(2\beta\left[RB(1)-RB(2)\right]\left[LB(1)-LB(2)\right]\right) \quad (2)$$

for the L-M direction. Here RA and RB are the expression levels of EphA and EphB receptors for axons 1 and 2, while LA and LB are the ligand densities of ephrinA and ephrinB in the corresponding target cells that are contacted by axons 1 and 2. The positive coefficients α and β are the parameters of the model; large values result in perfect mapping and zero values produce a random map. We use α=β =1000 throughout the paper. Signs in Eqs.(1-2) represent chemorepulsion between the EphA receptor and its ligands ephrinA and chemoattraction between the EphB receptor and its ligand ephrinB, so that $P_{EXCHANGE} < P_{RETAIN} = 1 - P_{EXCHANGE}$ for a pair of axons with the correct ordering. The discussion of the choice of probabilities (1-2) can be found in (Koulakov and Tsigankov, 2003).

## Results

The results of the simulation of temporal development for the wild-type case are shown in Fig.2. The system evolves from a random map towards forming an ordered map. We track in time tectal positions of axons originating from a circular region in the central retina. Note, that the axons initially positioned laterally to their correct TZ move up the gradient of ephrin-B1 (red arrow), while axons positioned medially drift down the gradient (blue arrow). The behavior of the latter may seem abnormal for our model because the model includes only attractive interaction in the EphB-ephrinB pair (see Model). What pushes the axons down the gradient of ephrin-B1? We demonstrate below that axonal competition provides the force that drives axons in the opposite direction to the one preferred by attractive interaction with ephrin-B1 density.

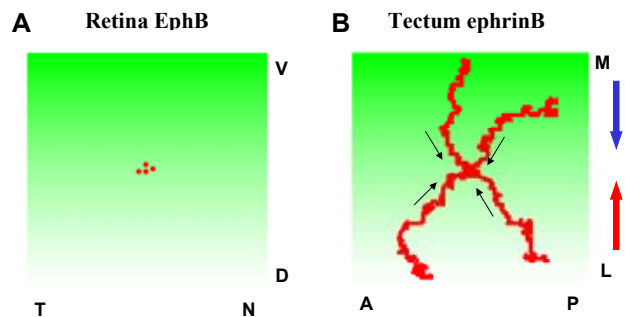

Figure 2 - Wild-type map development
Tectal trajectories (red lines, B) of axons originating in the central retina (red dots, A) are shown for a set of four axons. The temporal development is simulated for $5 \times 10^6$ iterations. The axons may drift up (red arrow) or down (blue arrow) the ephrin-B1 gradient. (T) temporal, (N) nasal, (V) ventral, (D) dorsal, (A) anterior, (P) posterior, (M) medial, (L) lateral.

To mimic studies in the transfected chick tectum (McLaughlin et al., 2003b), we add a constant level of ephrin-B1 density to a square region in central tectum. The results of the development are presented in Fig.3. To demonstrate the dynamics of axonal competition, we show concentration levels of axonal EphB receptor in tectum during the process of development. Axons with the highest EphB expression levels accumulate inside the ectopic region of high ephrin-B1 density because they experience the strongest attraction to this region. Axons with lower receptor expression are expelled from the ectopic zone by axonal competition. We track the position of a particular axon (red dot) that is initially within the ectopic zone and has a correct TZ (red empty circle) positioned medially from it. This axon drifts laterally (down the ligand gradient) away from its correct TZ, experiencing an effective repulsion from the regions of the high ephrin-B1 concentration. It is expelled further down the gradient during development by the growing area occupied by axons with the higher EphB expression levels (dark green area in Fig.3). These axons cannot be expelled themselves because their attraction to the ectopic region is the strongest. Thus, the regions with high ephrin-B1 density attract the 'strong' axons and they, in turn, through axonal competition, repel all other axons.

inhibited within the ectopic zone and the spatial distribution of their TZ is shaped by the edges of the ectopic domain. We compare this phenomenon to the behavior of passengers in subway. Although many passengers attempt to enter a newly arrived train, only limited number of them can, due to excluded volume interactions. The less motivated individuals left on the platform may seem to be repelled by the train, which is, of course, purely spurious. Thus, excluded volume interactions may generate effective repulsion by an attractive agent.

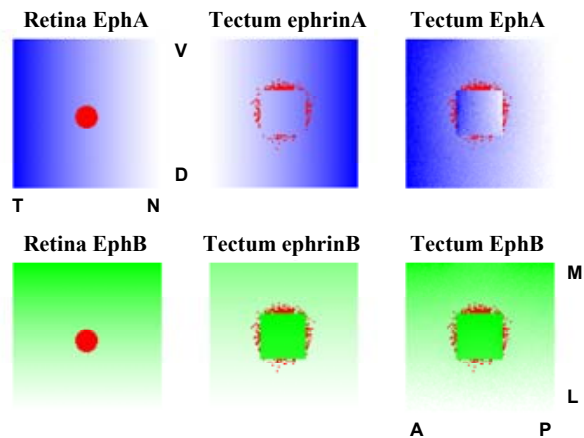

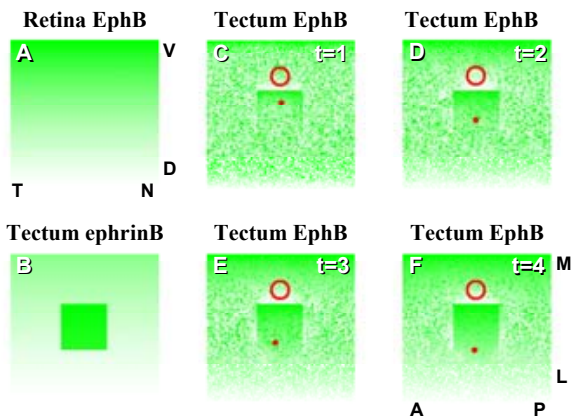

Figure 3 - Map development with the ectopic zone
Concentration profiles of EphB receptor in retina (A) and ephrin-B in tectum (B) are shown by the intensity of green. (C-F) Levels of EphB receptor expression of axons in tectum are shown at different stages of development. Temporal evolution of the map is shown for t =(1,2,3,4)x$10^6$ iterations. The red dot represents the position in tectum of a particular axon. The empty red circle shows the position of correct TZ for this axon. (T) temporal, (N) nasal, (V) ventral, (D) dorsal, (A) anterior, (P) posterior, (M) medial, (L) lateral.

The experimental study (McLaughlin et al., 2003b) also reports that ectopic domains of high ephrin-B1 density inhibit axonal arborization. Therefore, we analyze the TZ distributions of axons in the vicinity of the ectopic domain in the final map configuration. In Fig.4, we present positions of the axons originating from a circular region in the central retina. In the absence of the ectopic domain, these axons would project to the central tectum (see e.g. Fig.2). When the ectopic region of high ephrin-B1 density is coincident with the location of TZ, these axons are expelled from the ectopic domain by axons with higher EphB density due to axonal competition. Thus, the arborization of this set of axons is

Figure 4 - Avoidance of the ectopic domain
Positions of the axons originating from the circular region in central retina (left column) in tectum are shown for the final map configuration (right and central column) by red dots. The concentration profiles of the Eph receptors in the retina (left column), ephrins in tectum (central column) and axonal Eph expression levels in tectum are shown by the intensity of blue (EphA/ephrinA) and green (EphB/ephrinB). The final map configuration is shown after $10^7$ iterations. (T) temporal, (N) nasal, (V) ventral, (D) dorsal, (A) anterior, (P) posterior, (M) medial, (L) lateral.

## Discussion

Two distinct types of models are proposed to explain mechanism of retinotectal map formation (Prestige and Willshaw, 1975). One relies on the existence of two different interactions per axis between the RGC axons and the target cells. In mapping of the D-V retinal axis one interaction drives the axonal extensions in the medial direction in tectum and the second interaction makes the axons move laterally. The axonal TZs are formed where these interactions are in balance. These models are called 'dual gradient models' (group I according to (Prestige and Willshaw, 1975)) because two gradients of chemical labels in tectum provide these interactions. The models of the second type are based on axonal competition and only one type of interaction per axis with a gradient provided by target cells. In this case, axonal competition provides the repulsion needed to spread the axons along the whole tectum. Our model belongs to this type that is called the 'competition models' (group II in (Prestige and Willshaw, 1975)). We suggest here that in models of both types some axons are repelled by high ephrin-B1 density and that arborizations of these axons are inhibited by ectopic domains of high ligand density. Thus, we suggest that additional tests are needed to make a conclusion about repulsive interaction between EphB expressing axons and high ephrin-B1 density.

Some insight may come from whole eye cholera toxin injections. In dual gradient models all axons are repelled by high ephrin-B1 density. In contrast, in the competition model some axons are repelled and some are attracted by the ectpopic region of high ligand density. Therefore, in an experiment in which the retroviral transfection of tectum is combined with e.g. whole eye cholera toxin β subunit injection (Feldheim et al., 2000), dual gradient models predict a reduced density of axons within the ectopic domains, while no decrease in density would be detected in the competition model. Thus, if a decrease in axonal density within ectopic domains is observed, conclusions made by (McLaughlin et al., 2003b) of bifunctional ephrin-B1 action can be confirmed.

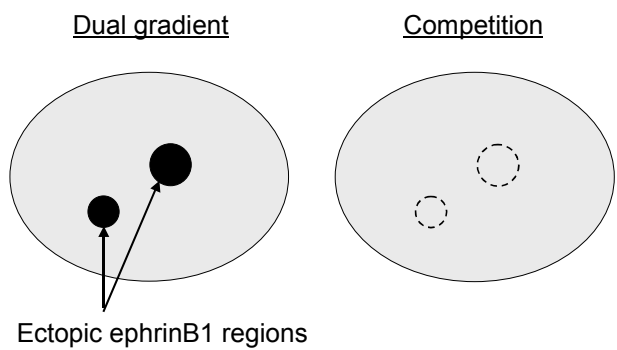

Figure 5- Whole eye cholera toxin injection experiment.
Whole eye cholera toxin injection may be used in transfected chick tectum to identify if the projections are made to the ectopic domains of high ephrin-B1 density (circular regions). Dual gradient model (left) predicts a reduced number of axons in the ectopic regions (black circles). Conversely, the competition model predicts that at least the same number of RGC axon projections will be formed within the ectopic domains (dashed empty circles).

We conclude that complex axonal behaviors may lead to experimental observations in transfected chick tectum (McLaughlin et al., 2003b). In particular, axonal competition combined with attractive interactions in the EphB-ephrinB pair may result in axonal repulsion from high ephrin-B1 density. We suggest an experimental test that may distinguish between the two different mechanisms of retinotectal map formation, one based on dual interactions with chemical labels and another based on axonal competition. We propose that it is difficult, if not impossible, to make qualitative inferences about attractive or repulsive interactions with chemical labels without the use of a particular quantitative model.